\documentclass[10pt, preprint]{aastex}

\usepackage[english]{babel}
\usepackage[utf8x]{inputenc}
\usepackage{amsmath}
\usepackage{graphicx}
\usepackage[colorinlistoftodos]{todonotes}

\newcommand{\frb}{FRB~121102}
\newcommand{\DMunits}{pc~cm$^{-3}$}
\newcommand{\sbw}{\Delta \nu_{\rm sb}}
\newcommand{\bw}{\Delta \nu}
\newcommand{\Nscint}{N_{\rm scint}}
\newcommand{\mId}{m_{I,d}}

\newcommand{\td}{\Delta t_d}
\newcommand{\tr}{\Delta t_r}
\newcommand{\nuG}{\nu_{\rm 1\,GHz}}
\newcommand{\Vkmps}{V_{\rm km/s}}

\newcommand{\change}{}

\shorttitle{High Frequency Bursts from FRB 121102}
\shortauthors{Spitler et al.}

\begin{document}
\title{Detection of Bursts from \frb\ with the Effelsberg 100-m Radio Telescope at 5 GHz and the Role of Scintillation}

\author{L.~G.~Spitler\altaffilmark{1},
W.~Herrmann\altaffilmark{2},
G.~C.~Bower\altaffilmark{3},
S.~Chatterjee\altaffilmark{4},
J.~M.~Cordes\altaffilmark{4},
J.~W.~T.~Hessels\altaffilmark{5,6},
M.~Kramer\altaffilmark{1,7},
D.~Michilli\altaffilmark{5,6},
P.~Scholz\altaffilmark{8},
A.~Seymour\altaffilmark{9},
A.~P.~V.~Siemion\altaffilmark{10,11,12}
 }
\affil{$^1$Max-Planck-Institut f\"ur Radioastronomie, Auf dem H\"{u}gel 69, D-53121 Bonn, Germany}
\affil{$^2$Astropeiler Stockert e.V., Astropeiler 2-4, D-53902 Bad M\"{u}nstereifel, Germany}
\affil{$^3$Academia Sinica Institute of Astronomy and Astrophysics, 645 N. A'ohoku Place, Hilo, Hawaii 96720, USA}
\affil{$^4$Cornell Center for Astrophysics and Planetary Science and Department of Astronomy, Cornell University, Ithaca, NY 14853, USA}
\affil{$^5$ASTRON, Netherlands Institute for Radio Astronomy, Postbus 2, 7990 AA, Dwingeloo, The Netherlands}
\affil{$^6$Anton Pannekoek Institute for Astronomy, University of Amsterdam, Science Park 904, 1098 XH Amsterdam, The Netherlands}
\affil{$^{7}$Jodrell Bank Centre for Astrophysics, University of Manchester, Alan Turing Building, Oxford Road, Manchester, M13 9PL, United Kingdom}
\affil{$^8$National Research Council of Canada, Herzberg Astronomy and Astrophysics, Dominion Radio Astrophysical Observatory, P.O. Box 248, Penticton, BC V2A 6J9, Canada }
\affil{$^9$National Astronomy and Ionosphere Center, Arecibo Observatory, PR 00612, USA}
\affil{$^{10}$Department of Astronomy and Radio Astronomy Lab, University of California, Berkeley, CA 94720, USA }
\affil{$^{11}$Radboud University, Nijmegen, Comeniuslaan 4, 6525 HP Nijmegen, The Netherlands }
\affil{$^{12}$SETI Institute, 189 N Bernardo Ave \#200, Mountain View, CA 94043, USA}

\begin{abstract}
\frb, the only repeating fast radio burst (FRB) known to date, was discovered at 1.4~GHz and shortly after the discovery of its repeating nature, detected up to 2.4~GHz \citep{ssh+16,ssh+16a}. {\change Here we present three bursts detected with the 100-m Effelsberg radio telescope at 4.85~GHz.} All three bursts exhibited frequency structure on broad and narrow frequency scales. Using an autocorrelation function analysis, we measured a characteristic bandwidth of the small-scale structure of 6.4$\pm$1.6~MHz, which is  consistent with the diffractive scintillation bandwidth for this line of sight through the Galactic interstellar medium (ISM) predicted by the NE2001 model \citep{cl02}. These were the only detections in a campaign totaling  22~hours in 10 observing epochs spanning five months. The observed burst detection rate within this observation was inconsistent with a Poisson process with a constant average occurrence rate; three bursts arrived in the final 0.3~hr of a 2~hr observation on 2016 August 20. We therefore observed a change in the rate of detectable bursts during this observation, and we argue that boosting by diffractive interstellar scintillations may have played a role in the detectability. Understanding whether changes in the detection rate of bursts from \frb\ observed at other radio frequencies and epochs are also a product of propagation effects, such as scintillation boosting by the Galactic ISM or plasma lensing in the host galaxy \citep{cwh+17}, or an intrinsic property of the burst emission will require further observations. 
\end{abstract}


\section{Introduction}
Fast radio bursts (FRBs) are millisecond-duration pulses of radio emission originating from so-far unidentified astrophysical sources \citep[e.g.][]{lbm+07,tsb+13}. All known FRBs have been discovered by surveys operating at either 800~MHz or 1400~MHz. The distances to the sources of FRBs, inferred from their large observed dispersion measures (DMs), range between $\sim$0.1 and 10~Gpc, which implies isotropic burst energies between $\sim10^{37}$ and $10^{40}$~erg \citep{pbj+16}. All but one of the known FRBs have been observed as one-off events and have positional precision too coarse to unambiguously identify a host galaxy, which is required for a direct distance measurement.  The one currently known exception is \frb.


\frb\ was discovered by the Arecibo Observatory in pulsar search data from the {\change Pulsar Arecibo L-band Feed Array (PALFA) survey} \citep{cfl+06,lbh+15} at a of DM = 557.4$\pm$2 \DMunits\  \citep{sch+14}. Ten additional bursts from \frb\ were detected in follow-up observations in 2015 May and June with Arecibo \citep{ssh+16}, making \frb\ the first and so-far only repeating FRB. Its repetitive nature enables extensive follow-up observations that are not possible with the one-off FRBs. Radio interferometric, optical imaging, and optical spectroscopic observations have shown that the bursting source is spatially coincident with a compact persistent radio source in a star formation region within a low-metallicity dwarf galaxy at a redshift of 0.193 \citep{clw+17,mph+17,tbc+17,bta+17}. {\change Neither persistent emission nor pulsed emission, coincident with radio bursts or otherwise,} has been detected in X-ray and gamma-ray observations \citep{sbh+17}. Also, the precise position allows for observations at higher radio frequencies, which are impractical for blind surveys because of the small instantaneous field of view. High S/N detections with Arecibo and the Green Bank Telescope (GBT) at radio frequencies between 4-8~GHz showed that \frb's bursts have complex time-frequency structure, are 100\% linearly polarized, and have an uncommonly large rotation measure  \citep[$\sim 10^5$ rad~m$^{-2}$,][]{msh+18,gsp+18}. Two models are currently favored for the \frb\ system: a young magnetar imbedded in a shell of ejecta or an energetic neutron star in the vicinity of a massive black hole.
{\change The host galaxy and \frb's location in a star formation region suggests a connection to hydrogen-poor superluminous supernovae and long gamma-ray bursts \citep[e.g.][]{tbc+17,mbm17}.  On the other hand, the luminosity of the persistent radio source suggests a low luminosity AGN, and the large and varying RM has similarities to the magnetar J1745-2900 in the Galactic center \citep{dep+18}. }

The spectra of bursts from \frb\ differ from the spectra typically observed in radio pulsars, which are well-modeled with a power law \citep[e.g.][]{kkg+03,jsk+18}. Broadband observations (0.3 to 8.4 GHz) of giant pulses from the Crab show that the spectra are consistent with a single power law about 70\% of the time, while the remaining 30\% show spectral flattening \citep{mat+16}.
Burst emission from \frb\ is poorly described by a power law; instead, emission occurs over a restricted range of bandwidth that is well-modeled with a Gaussian \citep[e.g.][]{lab+17}. \citet{ssh+16a} measured a characteristic bandwidth of the burst emission of 600~MHz for two of the GBT-detected bursts at 2~GHz, and \citet{lab+17} measured a characteristic bandwidth of 500~MHz in nine VLA-detected bursts at 3~GHz. 

Furthermore, the VLA observations of \frb\ were often accompanied by other radio telescopes, such as Arecibo and the 100-m Effelsberg radio telescope (hereafter simply Effelsberg), observing at 1.4~GHz and 4.85~GHz, respectively \citep{clw+17,lab+17}. Three of the nine VLA-detected bursts had simultaneous Arecibo coverage, and two of these also had simultaneous Effelsberg coverage. In one case a burst was detected simultaneously by the VLA and Arecibo but not at Effelsberg, while in the second case only the VLA detected the burst. The simultaneous detection between Arecibo and the VLA shows that some bursts are detectable over approximately a GHz in bandwidth, suggesting there is signifiant burst-to-burst variation in the bandwidth \citep{lab+17}. In summary, the radio emission from \frb\ occurs in confined ``islands" with a center frequency and bandwidth that changes from burst to burst. 

Previous observations of \frb\ at Arecibo and the VLA at lower radio frequencies suggested that the burst detection rate is time variable. On 2 June 2016 10 bursts were detected in 0.55 hours at Arecibo using the ALFA receiver and Mock spectrometers \citep{ssh+16}, whereas 13 epochs with Arecibo between 2015 November and 2016 January, each of $\sim$2 hours in duration, yielded only 1 burst \citep{ssh+16a}. Similarly, a $\sim$40 hour high-cadence campaign with the VLA in early 2016 yielded no detections, whereas a campaign in September 2016 with similar total observing time and observing cadence yielded nine detections \citep{clw+17,lab+17}. Clearly, detections of bursts from \frb\ are inconsistent with a constant Poisson detection rate. This is markedly different from the arrival times of giant pulses from the Crab pulsar, which are consistent with Poisson arrival times with a constant rate on a time scale of several hours \citep[e.g.][]{ksv10} with longer-term variations attributable to refractive interstellar scintillation \citep[e.g.][]{lcu+95}. 
Alternative statistical descriptions for \frb's burst detection rates have been proposed such as a Weibull distribution, which includes an additional parameter for clustering of the burst arrival times \citep{oyp18}. 

Investigating the detection rates at different radio frequencies will help disentangle whether the variable detection rate is intrinsic to the source or primarily due to propagation effects such as interstellar scintillations or plasma lensing \citep{cwh+17}. {\change For the typical instantaneous bandwidths of modern radio astronomy receivers, higher frequency observations ($\gtrsim 5$~GHz) of \frb\ will be more strongly influenced by scintillation boosting than observations at $\sim$1~GHz. Large gains from plasma lensing on the other hand are more likely to occur at lower radio frequencies for a given lens geometry \citep{cwh+17}. Also, the time scales for scintillation boosting are minutes to hours, while gain boosts from plasma lensing has been observed to change on timescales as short as 10~ms \citep{myc+18}. }

Understanding the broadband spectrum of \frb\ would help constrain emission and origin models. Therefore, we initiated multi-telescope simultaneous observations of \frb\ using the German international LOFAR stations (GLOW) at 150 MHz, Effelsberg at 4.85 GHz, and partial coverage with the Stockert 25-m radio telescope at 1.4 GHz. The results from the GLOW observations will be presented elsewhere. 

In Section~\ref{sec:oda} we will describe the observations and data analysis. The burst discoveries will be presented in Section~\ref{sec:discoveries}.  We discuss the role of scintillation in the detections in Section~\ref{sec:scint} and burst detection rates in Section~\ref{sec:rate}. A summary is given in Section~\ref{sec:conclusions}.
 
\section{Observations and Data Analysis}
\label{sec:oda}
\subsection{{\change Effelsberg 100-m Radio Telescope}}
Observations of \frb\ were conducted {\change between 2016 May 14 and 2016 September 18} with Effelsberg in the frequency band 4.6 to 5.1 GHz using the S60mm receiver, which has a system equivalent flux density (SEFD) of 18~Jy and a full-width half-max (FWHM) beam size of 2.4 arcmin at 4.85 GHz. Total intensity pulsar search mode data were recorded with the PSRIX pulsar timing backend \citep{lkg+16} with a time resolution of 51.2~$\mu$s and bandwidth of 500~MHz divided into 512 frequency channels, yielding a frequency resolution of $\Delta \nu \approx$ 0.98~MHz. For reference, the intrachannel DM smearing time for 560~\DMunits\ is $\sim$48~$\mu$s at 4.6 GHz. Note, no polarization information was preserved in these data, so a measurement of RM is not possible. 

When this observational campaign began, the uncertainty in the position of \frb\ was $\sim$3 arcmin \citep{ssh+16}. Because the FWHM beam size of Effelsberg at 4.85 GHz is only 2.4 arcmin, it was necessary to grid the Arecibo uncertainty region. Five grid positions were used: one directly on the best Arecibo position (POS\_A), two offset in declination by $\pm$1.2~arcmin (POS\_B, POS\_D), and two offset in right ascension by $\pm$1.2~arcmin (POS\_C, POS\_E), i.e. the centers of the offset beams are on the FWHM of the central beam position. The first VLA detection occurred on 2016 August 23, and thereafter we observed with a single pointing on the precise position (POS\_F). The names and positions of each pointing is given in Table~\ref{tab:pos}. 

\begin{deluxetable}{lcc}
\tablecolumns{3}
\tablecaption{Grid positions of Effelsberg 4.85~GHz observations\label{tab:pos} }

\tablehead{Name & R.A.\tablenotemark{a} & Decl.\tablenotemark{a} }
\startdata
POS\_A		&	05$^{\rm h}31^{\rm m}58^{\rm s}$	&	+33$^{\circ}08^{'}04^{''}$\\
POS\_B		&	05$^{\rm h}31^{\rm m}58^{\rm s}$	&	+33$^{\circ}09^{'}17^{''}$\\
POS\_C	&	05$^{\rm h}32^{\rm m}04^{\rm s}$	&	+33$^{\circ}08^{'}04^{''}$\\
POS\_D	&	05$^{\rm h}31^{\rm m}58^{\rm s}$	&	+33$^{\circ}06^{'}51^{''}$\\
POS\_E		&	05$^{\rm h}31^{\rm m}52^{\rm s}$	&	+33$^{\circ}08^{'}04^{''}$\\		
POS\_F		&	05$^{\rm h}31^{\rm m}  58.6^{\rm s}$ &	+33$^{\circ}08^{'}49.6^{''}$
\enddata
\tablenotetext{a}{Epoch J2000}
\end{deluxetable}

{\change The true position of \frb\ is only within the FWHM beam areas for POS\_F and the two grid pointings (POS\_A and POS\_B). The total amount of time spent observing \frb\ on one of these three positions was 22 hours over 10 observing epochs. The centers of POS\_A and POS\_B}  are offset from the true position by 46 and 27 arcseconds, respectively. For the observations from 16 September through 18 September, a 100-MHz bandpass filter centered on 4.85 GHz was mistakenly used, reducing the bandwidth by a factor of 5. {\change The minimum detectable flux density is increased by a factor of $\sqrt{5} \approx 2.24$ for these observations.} The total of 22 hours was spent on these three positions, and the exact start MJDs and durations of those observations are given in Table~\ref{tab:obs}. 

The PRESTO software package \citep{r01} was used to search for bursts. The radio frequency interference (RFI) package {\tt rfifind} was applied to the data, but this band is particularly clean and little data was flagged. Dedispersed time series were generated from 408~\DMunits\ to 712~\DMunits\ in steps of 8~\DMunits\ and downsampled by a factor of two. Candidate events were generated by convolving the dedispersed time series with boxcar templates with widths between a single sample and 20~ms and applying a signal-to-noise ratio (S/N) threshold of 5 using {\tt single\_pulse\_search.py}. The results were inspected manually for bursts.

\begin{deluxetable}{lccccc}
\tablecolumns{6}
\tablecaption{Dates, durations, grid positions, and number of bursts detected\label{tab:obs} in each observation}
\tablehead{UT Date & UTC time & MJD\tablenotemark{a} & Duration (s) & Pointing & $N_b$	}
\startdata
20160514		&	11:51:52.7	&	57522.49436	&	4320		& 	A	&	0	 \\ 
20160514		&	13:04:03.0	&	57522.54448 	&	4320		&	B	&	0	\\  
20160514		&	17:22:25.8	&	57522.72391 	&	1440		&	A	&	0	 \\ 
20160514		&	17:46:37.3	&	57522.74071 	&	1440		&	B	&	0	 \\ 
20160515		&	17:01:34.8	&	57523.70943 	&	1440		&	A	&	0	 \\ 
20160515		&	17:25:46.3	&	57523.72623 	&	1440		&	B	&	0	 \\ 
20160515		&	19:02:23.7	&	57523.79333 	&	1440		&	A	&	0	 \\ 
20160523		&	16:40:14.3	&	57531.69461 	&	1440		&	B	&	0	 \\ 
20160523		&	17:04:25.0	&	57531.71140 	&	1440		&	A	&	0	 \\ 
20160523		&	18:41:05.0	&	57531.77853 	&	1440		&	B	&	0	 \\ 
20160523		&	19:05:15.6	&	57531.79532 	&	1440		&	A	&	0	 \\ 
20160730		&	07:35:00.6	&	57599.31598 	&	1440		&	A	&	0	 \\ 
20160730		&	07:59:09.6	&	57599.33275 	&	1440		&	B	&	0	 \\ 
20160730		&	09:35:58.3	&	57599.39998 	&	1200		&	A	&	0	 \\ 
20160730		&	09:56:08.7	&	57599.41399 	&	1200		&	B	&	0	 \\ 
20160820		&	05:41:41.9	&	57620.23729 	&	1200		&	A	&	0	 \\ 
20160820		&	06:01:52.3	&	57620.25130 	&	1200		&	B	&	0	 \\ 
20160820		&	07:22:36.8	&	57620.30737 	&	1200		&	A	&	0	 \\ 
20160820		&	07:42:46.4	&	57620.32137 	&	1200		&	B	&	0	 \\ 
20160820		&	09:03:30.8	&	57620.37744 	&	1200		&	A	&	0	 \\ 
20160820		&	09:23:40.4	&	57620.39144 	&	1200		&	B	&	3	 \\ 
20160910		&	09:41:48.2	&	57641.40403 	&	5898		&	F	&	0	 \\ 
20160911		&	09:17:14.2	&	57642.38697 	&	10692	&	F	&	0	 \\ 
20160916\tablenotemark{b}	&	08:53:32.1	&	57647.37051 	&	10200	&	F	&	0	 \\ 
20160917\tablenotemark{b}	&	08:52:16.0	&	57648.36963 	&	9600		&	F	&	0	 \\ 
20160918\tablenotemark{b}	&	08:57:40.0	&	57649.37338 	&	9000		&	F	&	0	 \\ 
\enddata
\tablenotetext{a}{Topocentric times}
\tablenotetext{b}{Observing bandwidth limited to 100~MHz.}
\end{deluxetable}

\subsection{Stockert 25-m Radio Telescope}
\label{stockert}
Stockert is a 25-m radio telescope located near Bad M\"{u}nstereifel, Germany \citep{w07} and is operated by the Astropeiler Stockert e.V., {\change an amateur radio astronomy organization}. Observations were done with a dual-polarization, uncooled 1.4 GHz receiver, and the system has an SEFD of $\sim$1111 Jy. For reference, a 1~ms FRB would need a flux density of $>$25 Jy to yield S/N$>$10. High time resolution spectral data were recorded with 100~MHz of bandwidth centered on 1380 MHz and a time resolution of 218 $\mu$s. Stockert was pointed at the best, pre-localization position determined by Arecibo \citep{ssh+16}. The FWHM of the telescope beam is $\sim$30 arcmin, and the small positional error of 0.8 arcmin between the true position of \frb\ and the pointing position has a negligible effect on the sensitivity. 

\section{Burst discoveries}
\label{sec:discoveries}
A total of three bursts were detected with Effelsberg during this campaign, and throughout this paper we refer to them in chronological order as burst 1, burst 2, and burst 3, respectively. All three were detected in a single pointing (POS\_B) in a 0.2~hr window in the final 0.3~hr of the 2016 August 20 session. These were the earliest detections of \frb\ at a frequency $>$2.4~GHz. Details of the burst properties are described below and listed in Table~\ref{tab:bursts}.

\begin{deluxetable}{lcccccc}
\tablecolumns{5}
\tablecaption{Burst properties\label{tab:bursts} }
\tablehead{Name & S/N$_{\rm max}$ & $W_{t}$ (ms) & $W_{\nu}$ (MHz) & $S_{\rm max}$ (mJy) & TOA\tablenotemark{a} & UTC time\tablenotemark{a}}
\startdata
Burst 1 &    11   &   0.5 $\pm$ 0.1 & 350 &  300 $\pm$ 40 & 57620.392218422 & 09:24:47.672 \\
Burst 2 &      7   &   0.6 $\pm$ 0.1 & 250 &  200 $\pm$ 40 & 57620.394074525 & 09:27:28.039 \\
Burst 3 &      9   &   1.7 $\pm$ 0.3 & 400 &  100 $\pm$ 20 & 57620.399630199 & 09:35:28.049
\enddata
\tablenotetext{a}{Burst time of arrival referenced to the solar system barycenter and infinite frequency}
\end{deluxetable}

\subsection{Dispersion Measure}
{\change Because of the low S/N of the bursts and the small dispersive delay at higher radio frequencies, we do not fit for DM. Generically the uncertainties on a DM measurement depend on the width and S/N of a burst with narrow, strong bursts providing the most precise measurements. A burst with a large frequency-averaged S/N can be resolved into several frequency subbands, and DM can be measured with a precision corresponding to a dispersive delay across the band much shorter than the width of the burst. For the narrowest possible observed pulse duration in our data ($\sim$100~$\mu$s, see Section~\ref{sec:width}), this could theoretically yield a DM measurement with an uncertainty $\lesssim 1$\DMunits. 

In the case of our detections, the S/N is too low to generate subband profiles. Therefore, our only handle on DM is looking at S/N and burst width. The uncertainty on the measured widths is roughly $\sim$ 0.2~ms, and we assume that we could measure an increase in the width of roughly this order. The offset in DM from the true value corresponding to a dispersive delay comparable to 0.2~ms is 5~\DMunits. The uncertainty on such a fit is  larger than what can be measured at lower frequencies and is not constraining.} Therefore, throughout the analysis we assume a value of DM = 560~\DMunits. This choice is well-supported by the DM measurement of 559.7$\pm$0.1\DMunits\ from a particularly narrow and bright burst detected at a similar frequency with Arecibo four months later\citep{msh+18}.

\subsection{Burst durations and bandwidths}
\label{sec:width}
The FWHM pulse width was measured by fitting the band-averaged burst profile with a single Gaussian. The fitting was done using a least-squares-fitting technique with the Gaussian amplitude, position in time, width, and baseline as free parameters. The measured widths of the three bursts range from $\sim$0.5 to 1.7 ms (see Table~\ref{tab:bursts}). 

Generically, the measured pulse width ($W_t$) is estimated by $\sqrt{W_i^2 + t_{\rm samp}^2 + \Delta t_{\rm DM}^2 + \tau_{\rm s}^2}$ where $W_i$ is the intrinsic width of the burst, $t_{\rm samp}$ is the sampling time of the data, $ \Delta t_{\rm DM}$ is the intrachannel DM smearing time, and $\tau_{\rm s}$ is the pulse broadening timescale due to multi-path propagation effects \citep[e.g.][]{cm03}. The time resolution of the fitted profiles is $t_{\rm samp} = 97.65\, \mu$s. The intrachannel DM smearing for DM=560~\DMunits\ is $\sim 48\,\mu$s. 
The measured upper limit to pulse broadening in \frb\ is $<$1.5~ms at 1~GHz \citep{sch+14}, which scales to 14~$\mu$s at 4.85~GHz, i.e. much smaller than the time resolution of the data. The rms sum of the three instrumental and radio propagation factors is $\sim$0.1~ms, suggesting that we are temporally resolving the bursts but would not be able to resolve $\sim 10\mu$s structure as seen by \citet{msh+18}.

\begin{figure}
\centering
\includegraphics[scale=0.4]{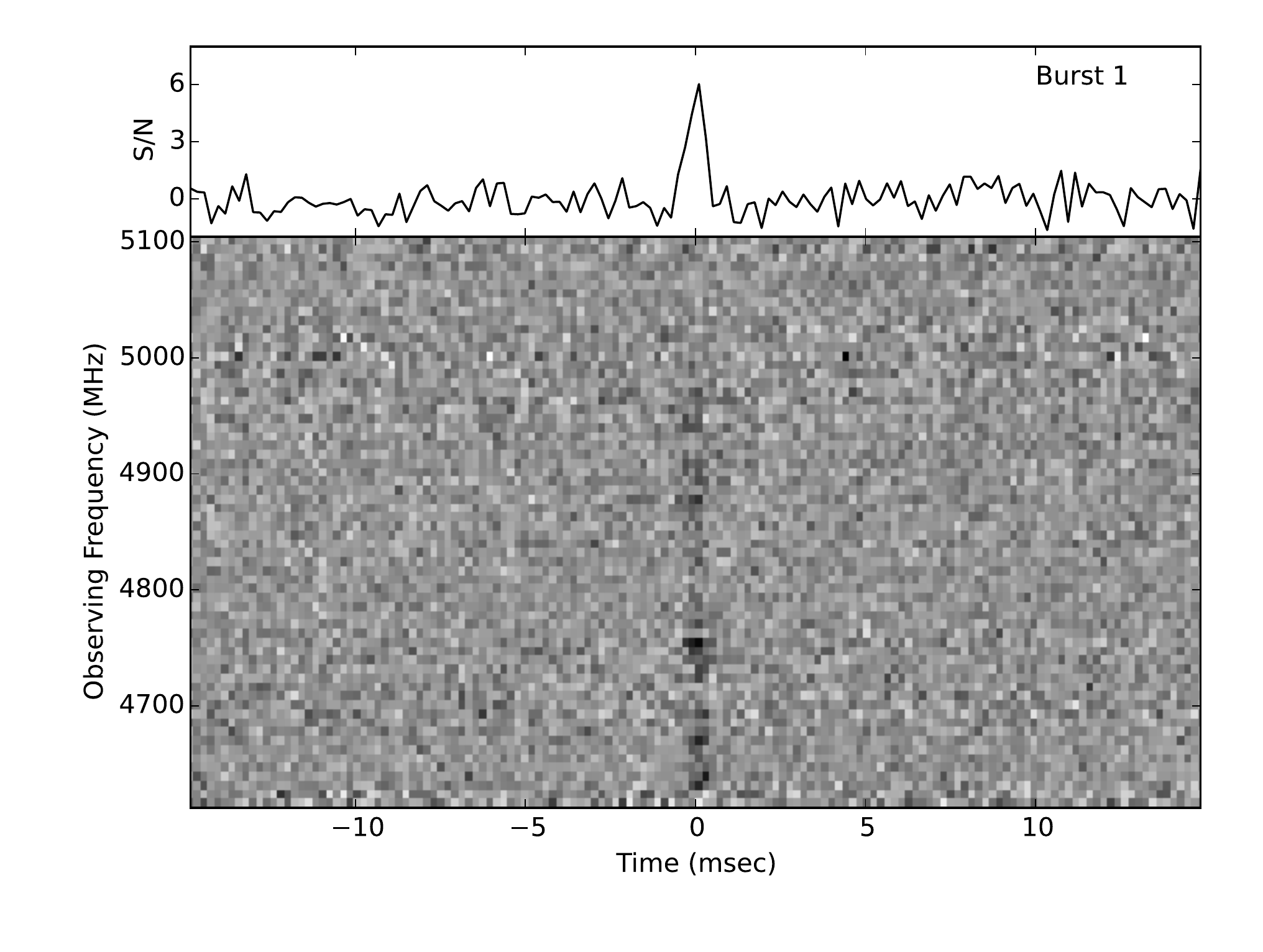}
\includegraphics[scale=0.4]{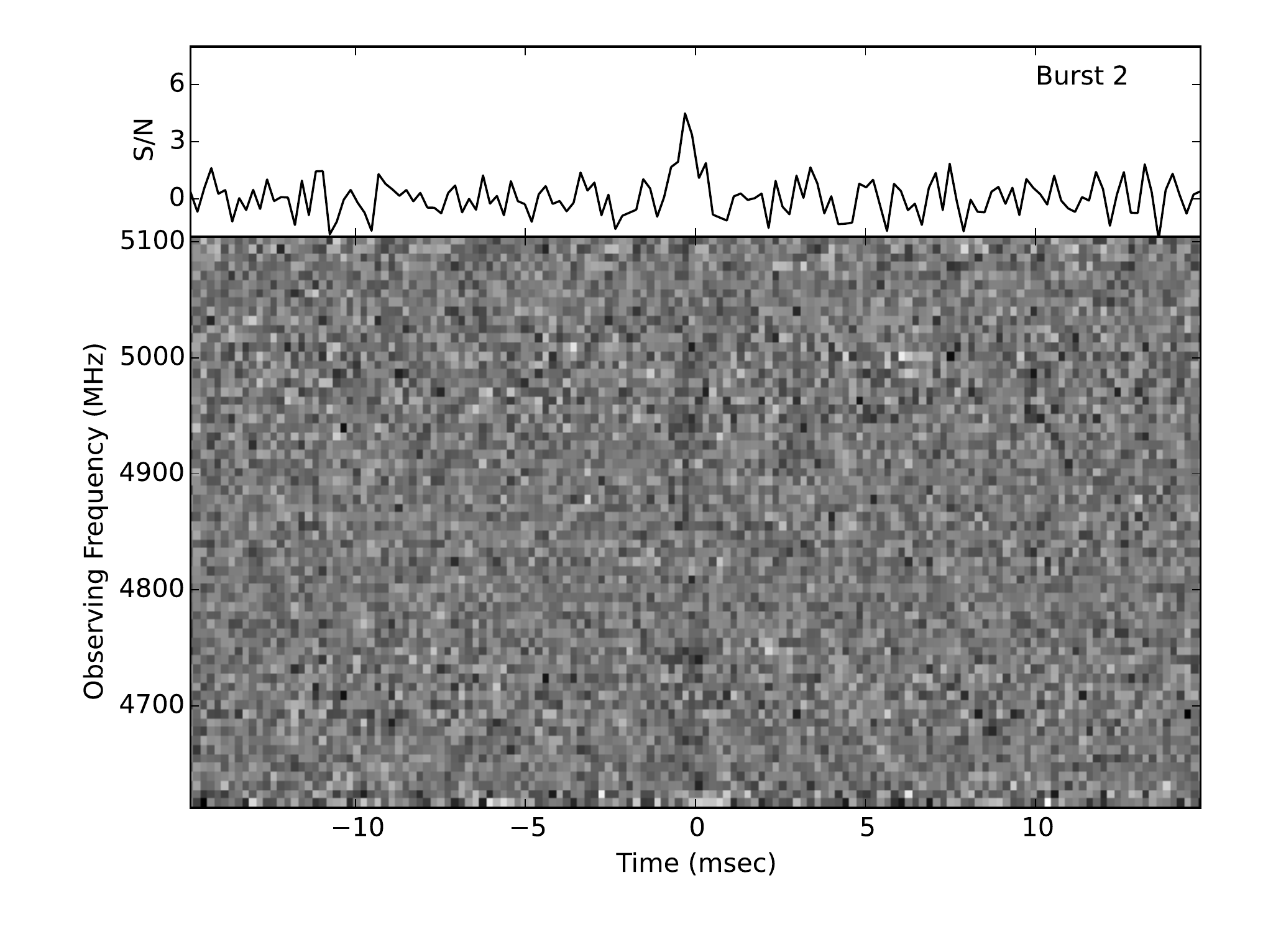}
\includegraphics[scale=0.4]{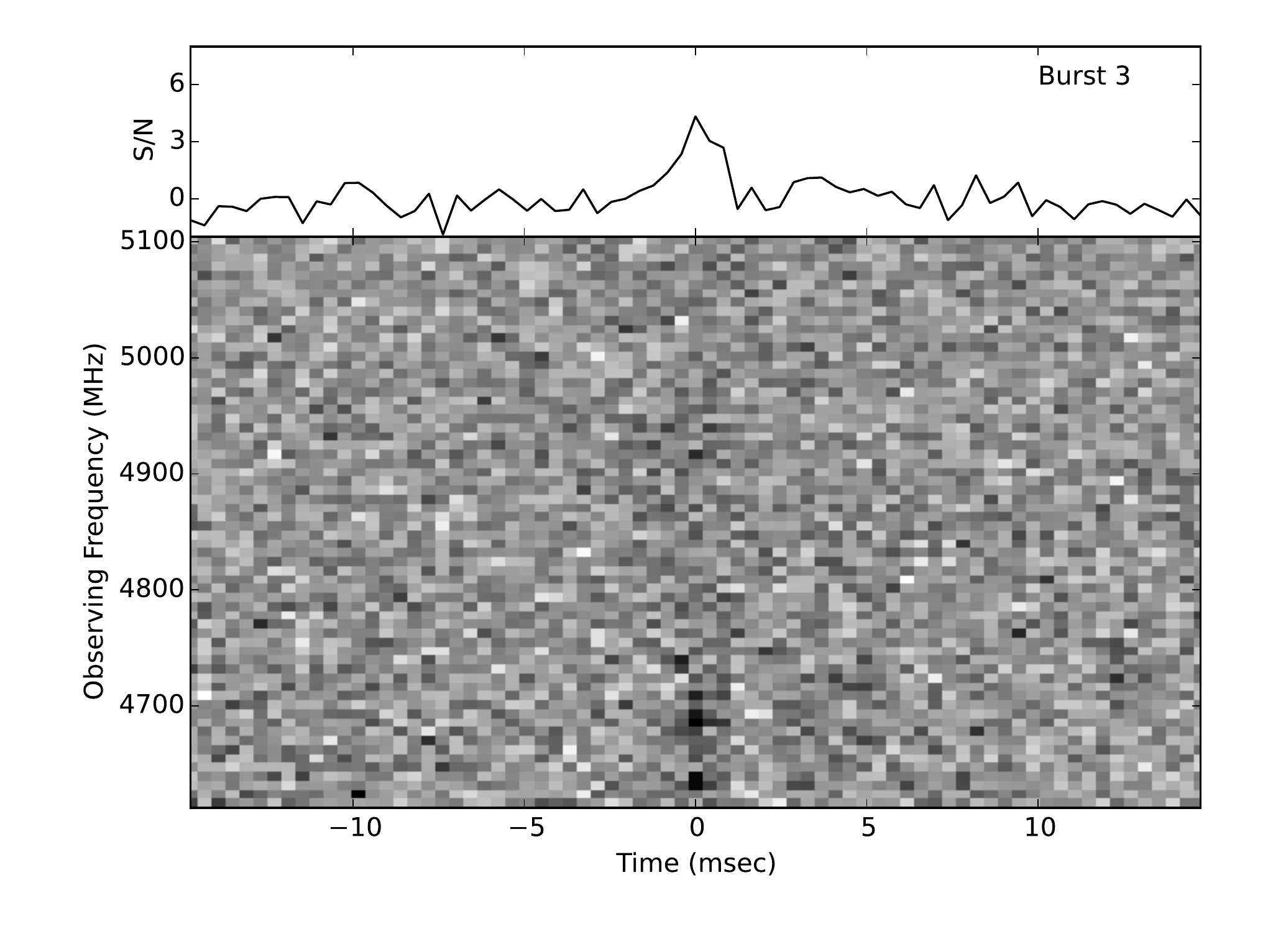}
 \caption{Dynamic spectra and average burst profiles of the three detected bursts in chronological order from top to bottom. Each burst has been dedispersed to a value of 560~\DMunits. The time resolution of the plots is 0.2, 0.2, and 0.4 ms, respectively, and the frequency resolution of the plots is  7.8 MHz. The S/Ns quoted in Table~\ref{tab:bursts} are higher than shown here, because those calculations assume the burst has been integrated in time into a single bin.
 }
\label{fig:dynspec}
\end{figure}

{\change It is clear from Figure~\ref{fig:dynspec} that the bursts' spectra are patchy. Furthermore, the fraction of the band containing flux may or may not be continuous, so characterizing the signal in terms of a bandwidth may or may not be appropriate. Instead, we use a spectral filling factor: $f_{\nu} = W_{\nu}/N_{\nu}$, where $W_{\nu}$ is the number of channels in the spectrum containing signal and  $N_{\nu}$ is the total number of frequency channels. We estimate the spectral filling factor using the spectral modulation index ($m_I$) \citep{sccs12}, which we describe briefly. 

If the intensity in each frequency channel of a burst's spectrum is given by $I(\nu)$, the spectral modulation index is defined as 
\begin{equation}
\label{eq:def}
	m_I^2 = \frac{\langle{I^2}\rangle - \langle{I}\rangle^2}{\langle{I}\rangle^2}, 
\end{equation}
where the brackets indicate averaging in frequency. The numerator of Equation~\ref{eq:def} is the variance of a burst's spectrum and the denominator is the square of the mean.  The spectral modulation index is therefore the normalized standard deviation of a burst's spectrum and is a metric for the ``broadbandedness" of a signal.

The two extreme cases are an intrinsically flat, broadband spectrum and an extremely narrowband spectrum. The spectral modulation index for the idealized broadband case is $\sqrt{N_{\nu}}/(S/N)$ and for the extreme narrowband case is $\sqrt{N_{\nu}}$. (Note, $S/N$ is the burst's single to noise ratio in the frequency-averaged time series.) The former case would be typical for single pulses from pulsars in the absence of measurable scintillation, whereas the latter case is typical for narrowband RFI. More generally a burst's spectrum could also have spectral structure, and $m_I$ would lie between the above limits. A more general expression parameterizing the modulation index is
\begin{equation}
\label{eq:mI}
m_I^2 = \frac{N_{\nu}}{\mbox{(S/N)}^2} + \frac{m_A^2}{f_{\nu}} + \frac{1-f_{\nu}}{f_{\nu}},
\end{equation}
where $m_A$ is the modulation index of the spectrum's signal \citep{sccs12}. For a scintillation-dominated spectrum, an intrinsic modulation index of $m_A \approx 1$ is appropriate. Note, for an intrinsically flat spectrum with $f_{\nu} =1 $ and $m_A = 0$, Equation~\ref{eq:mI} reduces to the idealized broadband case. 

Given a measured modulation index for each burst's spectrum and the assumption that $m_A = 1$, we can estimate $f_{\nu}$. The spectrum is defined as the average of the signal within the FWHM duration of the burst. The modulation indexes measured for these three bursts are 2.5, 3.7, and 2.8, respectively. According to Equation~\ref{eq:mI}, the corresponding filling factors are $f_{\nu} = $~0.7, 0.5, and 0.8, or in terms of frequencies ($W_{\nu} \Delta \nu$), 350 MHz, 250 MHz, and 400 MHz, respectively. }

The burst durations measured for the Effelsberg detections are consistent with those found for bursts detected at Arecibo at 4.5~GHz \citep{msh+18} and at the GBT at 4-8 GHz \citep{gsp+18}. In many cases the AO and GBT sample show multiple sub-bursts. Burst 3 has a width roughly twice that of bursts 1 and 2, which may suggest multiple sub-bursts, but the S/N is too low to make a definitive conclusion. Overall the sample of bursts above 4~GHz shows that the typical burst durations are shorter than at 1.4 GHz. The spectral properties we observed in the Effelsberg bursts are also broadly consistent with the Arecibo and GBT sample, which show burst-to-burst variations in bandwidth and structure on two frequency scales.   

\subsection{Flux density}
We use the radiometer equation to estimate the integrated flux density of the burst detections: 
\begin{equation}
S_{\nu} = \mbox{S/N} \frac{\mbox{SEFD}} {\sqrt{N_p \Delta \nu W_t}},
\end{equation}
where $S_{\nu}$ is the flux density and $N_p$ is the number of polarizations. The maximum S/N was estimated by averaging over the bins contained within the FWHM pulse width determined by the fitting described in Section~\ref{sec:width}. This corresponds to the S/N obtained from ideal matched filtering in which all the flux lands in a single time bin. The flux densities of the three bursts are given in Table~\ref{tab:bursts} for $N_p = 2$ and the measured $W_t$. The uncertainties on the flux densities correspond to the rms noise for each of the bursts, again based on the radiometer equation. 

We need to correct the SEFD to account for the off-axis detections. We define $\eta$, which is unity for an on-axis detection and $<1$ for an off-axis detection, and estimate it  by modeling the Effelsberg beam with a simple Gaussian with a FWHM of 146 arcseconds at 4.85 GHz. The burst detections occurred with the beam center pointed at POS\_B, which is offset from the best known position, POS\_F, by 27~arcseconds, yielding $\eta = 0.9$, which gives SEFD$_{c}$ = SEFD/$\eta$ = 20~Jy. For reference, $\eta = 0.75$ for pointing position POS\_A. Note, $\eta$ is frequency dependent and can lead to a significant instrumental spectral index \citep[e.g.][]{sch+14}. But in this case,  $\eta$ varies by only 2\% at the top and bottom of the bands due to the small fractional bandwidth, so we simply use the value at the band center. 

Note, the sessions in 2016 September had partial simultaneous coverage with Arecibo and the VLA. The limits on a broadband spectrum from these observations were discussed in detail in \citet{lab+17}. 

\subsection{Constraints on broadband spectrum}
{\change The observations with the Stockert telescope on 20 August 2016  began at 05:45 UTC, continued until around 11:30 UTC, resumed at 12:57 UTC, and   ended at 13:53 UTC for a total of 6.3 hrs  on source. The first observation block overlapped entirely with the Effelsberg observation.} No bursts from \frb\ were detected with the Stockert telescope. A blind search for bursts was done, in addition to a manual inspection of the raw data at the expected arrival time of each burst at 1.4 GHz after accounting for the dispersive delay. As mentioned in Section~\ref{stockert}, the estimated minimum detectable flux density for a 1~ms burst is roughly 25~Jy. {\change Note, the flux densities of the 5 GHz Effelsberg detections are two orders of magnitude lower than this.}

Previous observations show that the spectrum of \frb\ is characterized by ``islands" of emission. In order for a burst to be detected simultaneously at 1.4 GHz and 4.85 GHz by Stockert and Effelsberg, the peak of the island would likely need to be near 1.4 GHz and have a characteristic bandwidth of $\sim$2~GHz to include the Effelsberg band. The single, simultaneous VLA-Arecibo detection does show that bursts with characteristic bandwidths greater than approximately 1 GHz do occur, but perhaps only rarely. 

It is also possible that bursts from \frb\ are also simply too faint for Stockert to detect, but there has been one claimed detection of a burst with a peak flux density of 24$\pm$7~Jy from \frb\ \citep{ola+17}. {\change The brightest VLA detection occurred 14 days after this observation and had a flux density of $\sim$3 Jy. 
If the detection rate was roughly constant between 2016 August 20 and 2016 September 22 (the end of the VLA campaign), than a rough rate of $\sim$1~Jy bursts is one every 40 hours. Assuming the statistical distribution of burst flux densities has no frequency dependence, the probability of detecting a 25~Jy burst in 6.3 hours is small. Therefore, these observations are not constraining for the broadband spectrum of \frb.}

\section{Diffractive Interstellar Scintillation}
\label{sec:scint}
All measurements or constraints on multi-path propagation effects in the properties of \frb's bursts are consistent with being imparted by the Galactic interstellar medium (ISM). For example, the angular size of the persistent radio counterpart is consistent with the angular broadening size predicted by NE2001 \citep{mph+17}.
The diffractive scintillation bandwidth predicted by NE2001 for the line of sight to \frb\ is 
\begin{equation}
\sbw \sim 8\; \mbox{kHz}\; \nu_{\rm GHz}^{\beta},
\end{equation}
where $\nu_{\rm GHz}$ is the radio frequency in GHz \citep{cl02}. 
In simple models for the distribution of density fluctuations (such as uniform phase screens or a uniform 3D medium), an index $\beta = 4$ is expected if density fluctuations follow a power-law spectrum with the spatial wavenumber scaling as $\lambda^{-4}$ or if the medium has a single scale size.   A Kolmogorov spectrum with an inner scale also gives $\beta = 4$ if the scattering is dominated by fluctuations at this scale, which occurs in some cases.  In others, the Kolmogorov spectrum typically gives $\beta$ = 22/5 = 4.4.  More complex spatial distributions (e.g. irregular screens) can alter these scaling laws substantially \citep{cl01}.

We estimate the diffractive scintillation bandwidth using an autocorrelation function (ACF) analysis \citep[e.g.][]{cwb85}. Only burst 1 had sufficient S/N to characterize $\sbw$. First, a short segment of data was extracted around burst 1 with full frequency and time resolution. The data were bandpass corrected by normalizing by the median bandpass. The PSRIX spectrometers impose a highly scalloped spectrum on the data. The bandwidth of each scallop is $\sim$15.6 MHz, and the power on the edges of the scallop is half the peak. Below we show that this does not adversely affect our ability to characterize the diffractive scintillation bandwidth. 

An average burst spectrum was calculated by summing in time over the samples that are within the burst FWHM. The ACF of the burst is then calculated with the following normalization:
\begin{equation}
A(\delta \nu) = \frac{1}{\sigma_{I}^2}\sum_{\nu} [I(\nu + \delta \nu) - \bar{I}][I(\nu) - \bar{I}],
\end{equation}
where $I(\nu)$ is the total intensity in frequency channel $\nu$ and $\bar{I}$ and $\sigma_{I}^2$ are the mean and variance of the spectrum \citep{cwb85}.
The diffractive scintillation bandwidth is measured by fitting the ACF about the zero lag with a Gaussian function of the form $f(\delta \nu) = A e^{{-\ln(2)(\delta \nu/\sbw)}^2}$, where $A$ is the amplitude and $\sbw$ is the half width at half maximum, using a least squares routine. The zero-lag noise spike was excluded from the fit.

\begin{figure}
\centering
\includegraphics[scale=0.4]{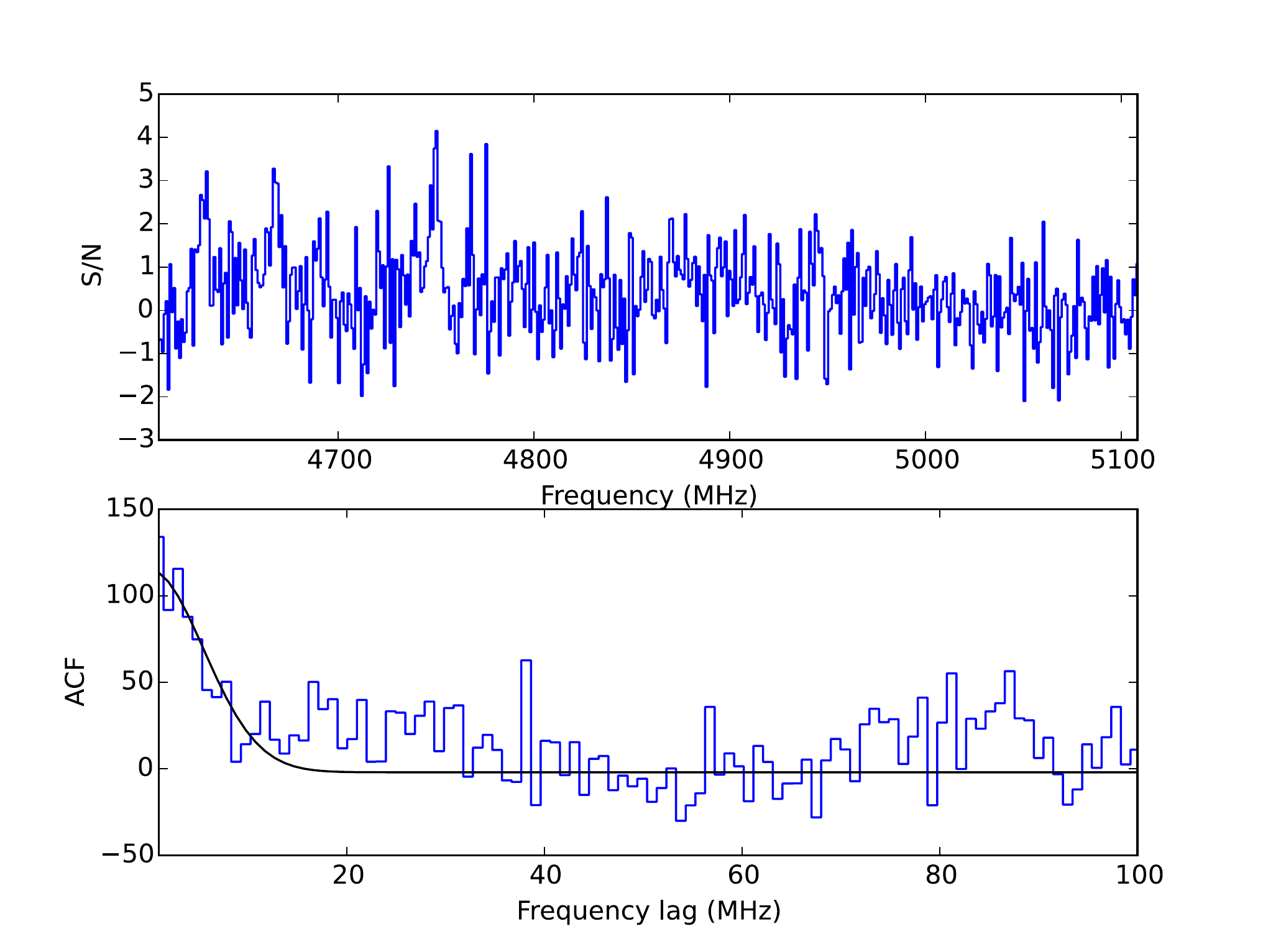}
 \caption{The spectrum (top) and autocorrelation function (below) of Burst 1. The best-fit Gaussian is overplotted on the calculated ACF.
}
\label{fig:ACF}
\end{figure}

The spectrum and ACF overlaid with the best-fit Gaussian for burst 1 are shown in Figure~\ref{fig:ACF}. We estimated the scintillation from the full-bandwidth spectrum, as well as the bottom half of the spectrum, which contains most of the signal. The formal fits and 1-$\sigma$ uncertainties are $\sbw = 6.5\pm0.9$~MHz and $4.0\pm0.7$~MHz for the full band and lower half, respectively. Note, these uncertainties are the formal, statistical uncertainty and do not reflect the estimation uncertainty from having a finite number of scintles ($\Nscint$) across the band. The fractional estimation uncertainty is $1/\sqrt{N_{\rm scint}} \sim 1/\sqrt{0.3 \bw/\sbw}$, where the factor of 0.3 accounts for the low filling factor of the scintles. The estimation uncertainty for these two ACFs is 1.4 MHz and 0.9 MHz, respectively. Taking the root-mean-square sum of the statistical uncertainty and estimation uncertainty we get the final $\sbw$ estimates of $6.5 \pm 1.6$~MHz and $4.0 \pm 1.2 $~MHz, for the full and half-bandwidth estimates, respectively. By comparison, \citet{msh+18} measured a diffractive scintillation bandwidth of 2-5 MHz for the bursts detected at Arecibo at 4.5 GHz. The measured $\sbw$ values are broadly consistent with the estimates from the NE2001, further suggesting that the scattering and scintillation properties of \frb\ are dominated by the ISM in our Galaxy.  

To explore the impact of the highly scalloped bandpass on our fits, we injected Gaussian-shaped ``scintles" into a real, off-pulse bandpass, as well as a simulated flat, white noise bandpass. The injected scintles had identical diffractive scintillation bandwidths but with central positions distributed with uniform probability within the band. The diffractive scintillation bandwidth was then measured using the method described above. This process was repeated 40 times for a range of simulated diffractive scintillation bandwidths and for the real and simulated flat bandpass. For bandwidths narrower than the width of the spectral scalloping ($\sim$ 16~MHz), the differences in the medians of the distributions of the measured $\sbw$ from real and flat bandpass were smaller than the 1-$\sigma$ width of the distributions in $\sbw$ from each bandpass type individually. Since the measured $\sbw$ in our data were much narrower than the scalloping bandwidth, we conclude that the measurements were not adversely affected by the bandpass shape. 

\section{Burst detection rate}
\label{sec:rate}
The total observing time {\change during this campaign} with Effelsberg at 4.85 GHz on positions POS\_A, POS\_B, or POS\_F is 22 hours.  On 2016 August 20 only two of the five grid pointings (POS\_A \& POS\_B) covered the true position of \frb. POS\_B directly followed POS\_A, and this pair was observed three times throughout the observation session. The duration of each pointing position was 0.33~hr, and the separation in time between the start of subsequent observations of the same grid position was 1.67~hr. The total time on source was therefore two hours spanning a four hour time span. {\change Also, the bursts were detected in the final 0.3~hr of the observation and is therefore a lower limit to the span of time during which bursts could have been detected. }

The average burst detection rate assuming the full 22 hours of observations is $0.1\pm0.05$ bursts hr$^{-1}$ above a peak flux density threshold of 0.1~Jy (assuming S/N = 5, $W_t$ = 1~ms, and $W_{\nu}$ = 500~MHz). {\change Note, the equivalent fluence limit is 0.1~Jy~ms for the assumed 1~ms burst duration.} The uncertainties assume Poisson statistics and a 68\% confidence interval. It is now well-established that the burst detection rate of bursts from \frb\ is variable on time scales longer than a few hours. If we instead assume that the detection rate is constant over the 2016 August 20 observing epoch, the average detection rate in these two hours is $1.5\pm1.0$~bursts hr$^{-1}$. Again, this may be an invalid assumption given that the bursts were detected in a $\sim$0.2~hr window. We tested the hypothesis that the arrival times of the bursts are consistent with arrival times randomly distributed with the two hours of on-source time, which would be expected for a Poisson process, using the Kolmogorov-Smirnov (KS) test. We can reject this hypothesis at the 99.5\% level. 
Therefore, we observed a change in the detection rate of bursts from \frb\ during the 2016 August 20 observation. 
Interestingly, \citet{gsp+18} observed a similar phenomena one year later; 21 bursts were detected in the first hour of observations and none in the following four hours. 

It should be noted, the sole epoch with Effelsberg 4.85 GHz detections occurred three days before the first VLA detection, i.e.\ during the several week period of time when the observed burst detection rate was higher than average. Furthermore, the Effelsberg observations are concentrated in 2016 May and 2016 August and September, overlapping with the two VLA campaigns at 3 GHz. Therefore, we now compare our observed rates to the VLA rates. \citet{lab+17} measure an average burst detection rate above a fluence of 0.2~Jy~ms of 0.16$\pm$0.05~hr$^{-1}$ from all observations in the spring and fall 2016 campaigns and a higher rate of 0.3$\pm$0.1~hr$^{-1}$ from all observations during the fall 2016 campaign. The rate of Effelsberg detections from all observations in 2016 August and September is 0.2$\pm$0.1~hr$^{-1}$. Note, the fluence threshold for the VLA and Effelsberg detections are similar, and the rates are consistent at the 1$\sigma$ level. This may suggest that the spectrum of \frb\ is roughly flat between these frequencies, although we argue below that propagation effects may have played a role in these detections. As such, it is not possible to make definitive statements about the intrinsic spectrum of \frb\ from the 4.85 GHz rate.  

The apparent burst clustering could be an intrinsic property of \frb\ or due to extrinsic effects such as scintillation from multi-path propagation \citep[e.g.][]{cr98} or plasma lensing \citep{cwh+17}. Given that the S/N's of these bursts are low, it is possible that the flux of the bursts was boosted above the detection threshold by a combination of diffractive or refractive scintillation. For strong scattering the intensity of a point source can be 100\% modulated if $\Nscint \sim 1$, i.e. when the bandpass is covered by a single scintle ($\sbw \sim \bw$). The modulation index in this case is $\mId = 1$.  If instead $\Nscint > 1$, the modulation is reduced by $f_B \sim 1/\sqrt{1+0.3\bw/\sbw}$. Taking as a characteristic value the average of the measured values in Section~\ref{sec:scint}, i.e. $\sbw \sim$ 5.4 MHz, the modulation index reduces to $\mId \approx 0.2$. Given that our detections were all close to the detection threshold, it is plausible that intensity scintillations played a role in the detectability of these three bursts. 

The temporal variations in the intensity can be characterized by diffractive and refractive timescales, which depend on the line of sight through the Galaxy, the observing frequency, and the perpendicular velocity of the Earth or source relative to the ISM. Here we use the NE2001 model to estimate the timescales for the line of sight to \frb.  The refractive timescale for this line of sight is $\tr = 21,500 \Vkmps \nuG^{-2.2}$ days \citep{ssh+16}. In the case of Galactic pulsars, the {\change pulsar's} motion typically dominates, and $\Vkmps$ is $\sim 100$~km~s$^{-1}$. Given its significantly larger distance, the proper motion of \frb\ is negligible, and the dominant effect is the motion of the ISM relative to the Earth, which is slower than typical pulsar velocities. Here we assume $\Vkmps \sim $1 to 10~km~s$^{-1}$. For an observing frequency of 4.85 GHz, this yields a time scale of $\tr \approx$~67 to 670 days, i.e. a couple of months to a couple of years. Therefore, refractive scintillations are not likely {\change to} cause a sudden increase in the observed burst rate, but may have contributed to an average increase in the observed signal flux and could influence detection rates in long term monitoring.  

For diffractive scintillations, the time scale for \frb's line of sight is $\td = 15,000 \nuG^{-1} \Vkmps^{-1}$~s, where $\nuG$ is the observing frequency in GHz and $\Vkmps$ is the velocity in the plane of the sky in km~s$^{-1}$ \citep{ssh+16}. Again assuming  $\Vkmps \sim $1 to 10~km~s$^{-1}$ the diffractive scintillation timescale at 4.85~GHz is $\sim$3000 to 300~s, i.e. a few minutes to roughly an hour. The time scale over which we saw the emission turn on was on the order of thousands of seconds, which is broadly consistent with the time scales predicted from diffractive scintillations in the Milky Way. 

\section{Discussion and Conclusions}
\label{sec:conclusions}
Recently, \citet{ffb+18} presented the first FRB that could be coherently dedispersed at discovery {\change (FRB~170827)} and is therefore the only other FRB besides \frb\ whose profile is not dominated by intrachannel DM smearing or scattering. FRB~170827 was discovered at 835~MHz by the UTMOST survey and has three sub-bursts and clear frequency structure {\change on two scales, one of which is attributed to diffractive scintillation.} High signal-to-noise ratio (S/N) bursts from \frb\ with coherent dedispersion from observations at Arecibo and the GBT have also shown complex time-frequency structure with multiple sub-bursts \citep[][Hessels et al., in prep]{msh+18,gsp+18}, and the qualitative similarities between \frb\ and FRB~170827 are notable. Therefore, the biggest observational differences between \frb\ and the rest of the FRB population is arguably the repeatability and large {\change measured} RM, which is at least 400 times larger than the FRB with the next largest RM. (Note, the data for FRB~170827 includes only a single hand of circular polarization, so no RM measurement can be made.)

Disentangling how much of the observed variability in both time and frequency is intrinsic to the source and how much is due to plasma propagation effects in \frb's host galaxy and in our Galaxy will require a long-term campaign of simultaneous or semi-simultaneous multi-frequency radio observations, ideally with wide instantaneous bandwidths. A key observable is burst detection rates measured at different radio frequencies. Galactic diffractive interstellar scintillation varies on a {\change timescale} that scales as $\nu^{-1}$. By comparing the timescales of variability in the observed detection rates at higher and lower frequencies, one could verify that the variations are consistent with Galactic scintillations. 

{\change On the other hand, the plasma lensing magnifies over a limited frequency range. A burst with an intrinsically broadband spectrum would be observed as having a spectrum with islands of emission \citep{cwh+17}}. This could explain a high rate of burst detections at one frequency and a significantly lower rate at higher and lower frequencies.  {\change The constraints on the physical parameters of the medium near \frb\ based on its DM and RM are consistent with would what be required for plasma lensing to occur \citep[][Hessels et al., in prep]{msh+18}.  Recently \citet{myc+18} observed strong lensing in single pulses from a black widow pulsar when the pulsar passed behind the stellar wind of its main sequence companion. The dynamic spectra of the pulses are remarkably similar to bursts from \frb. If the environment near \frb\ is more favorable to plasma lensing than the environments of the other FRBs, it would help explain why this source is observed as a repeating source, while the other FRBs are not \citep{cwh+17}.}  

{\change Here we presented the detection of bursts from \frb\ at a 4.85 GHz.} We detected three bursts with the Effelsberg radio telescope at 4.85~GHz. Our high-frequency search comprised a total of $\sim$22 hrs of observing in ten epochs spanning 2016 May through September. The three detections occurred in the last 0.3~hr of a two-hour observing session on 2016 August 20. This is inconsistent with a Poisson distribution with a constant rate within the observation. We also measure a diffractive scintillation bandwidth in the spectrum of the brightest burst that is fully consistent with what is expected from our Galaxy for that line of sight. Estimations of the scintillation timescale and modulation index suggest that intensity scintillations may be playing a role in the detectability of these bursts. 

\acknowledgments
Based on observations with the 100-m telescope of the MPIfR (Max-Planck-Institut f\"{u}r Radioastronomie) at Effelsberg. L.G.S. acknowledges financial support from the ERC Starting Grant BEACON under contract number 279702, as well as the Max Planck Society. S.C. and J.M.C. acknowledge support from the NANOGrav Physics Frontiers Center, funded by the National Science Foundation award number 1430284. J.W.T.H. is a Netherlands Organization for Scientific Research (NWO) Vidi Fellow and, together with D.M. acknowledges funding for this work from ERC Starting Grant DRAGNET under contract number 337062. P.S. holds a Covington Fellowship at DRAO.


\begin{thebibliography}{37}
\expandafter\ifx\csname natexlab\endcsname\relax\def\natexlab#1{#1}\fi

\bibitem[{{Bassa} {et~al.}(2017){Bassa}, {Tendulkar}, {Adams}, {Maddox},
  {Bogdanov}, {Bower}, {Burke-Spolaor}, {Butler}, {Chatterjee}, {Cordes},
  {Hessels}, {Kaspi}, {Law}, {Marcote}, {Paragi}, {Ransom}, {Scholz},
  {Spitler}, \& {van Langevelde}}]{bta+17}
{Bassa}, C.~G., {Tendulkar}, S.~P., {Adams}, E.~A.~K., {Maddox}, N.,
  {Bogdanov}, S., {Bower}, G.~C., {Burke-Spolaor}, S., {Butler}, B.~J.,
  {Chatterjee}, S., {Cordes}, J.~M., {Hessels}, J.~W.~T., {Kaspi}, V.~M.,
  {Law}, C.~J., {Marcote}, B., {Paragi}, Z., {Ransom}, S.~M., {Scholz}, P.,
  {Spitler}, L.~G., \& {van Langevelde}, H.~J. 2017, \apjl, 843, L8

\bibitem[{{Chatterjee} {et~al.}(2017){Chatterjee}, {Law}, {Wharton},
  {Burke-Spolaor}, {Hessels}, {Bower}, {Cordes}, {Tendulkar}, {Bassa},
  {Demorest}, {Butler}, {Seymour}, {Scholz}, {Abruzzo}, {Bogdanov}, {Kaspi},
  {Keimpema}, {Lazio}, {Marcote}, {McLaughlin}, {Paragi}, {Ransom}, {Rupen},
  {Spitler}, \& {van Langevelde}}]{clw+17}
{Chatterjee}, S., {Law}, C.~J., {Wharton}, R.~S., {Burke-Spolaor}, S.,
  {Hessels}, J.~W.~T., {Bower}, G.~C., {Cordes}, J.~M., {Tendulkar}, S.~P.,
  {Bassa}, C.~G., {Demorest}, P., {Butler}, B.~J., {Seymour}, A., {Scholz}, P.,
  {Abruzzo}, M.~W., {Bogdanov}, S., {Kaspi}, V.~M., {Keimpema}, A., {Lazio},
  T.~J.~W., {Marcote}, B., {McLaughlin}, M.~A., {Paragi}, Z., {Ransom}, S.~M.,
  {Rupen}, M., {Spitler}, L.~G., \& {van Langevelde}, H.~J. 2017, \nat, 541, 58

\bibitem[{{Cordes} {et~al.}(2006){Cordes}, {Freire}, {Lorimer}, {Camilo},
  {Champion}, {Nice}, {Ramachandran}, {Hessels}, {Vlemmings}, {van Leeuwen},
  {Ransom}, {Bhat}, {Arzoumanian}, {McLaughlin}, {Kaspi}, {Kasian}, {Deneva},
  {Reid}, {Chatterjee}, {Han}, {Backer}, {Stairs}, {Deshpande}, \&
  {Faucher-Gigu{\`e}re}}]{cfl+06}
{Cordes}, J.~M., {Freire}, P.~C.~C., {Lorimer}, D.~R., {Camilo}, F.,
  {Champion}, D.~J., {Nice}, D.~J., {Ramachandran}, R., {Hessels}, J.~W.~T.,
  {Vlemmings}, W., {van Leeuwen}, J., {Ransom}, S.~M., {Bhat}, N.~D.~R.,
  {Arzoumanian}, Z., {McLaughlin}, M.~A., {Kaspi}, V.~M., {Kasian}, L.,
  {Deneva}, J.~S., {Reid}, B., {Chatterjee}, S., {Han}, J.~L., {Backer}, D.~C.,
  {Stairs}, I.~H., {Deshpande}, A.~A., \& {Faucher-Gigu{\`e}re}, C.-A. 2006,
  \apj, 637, 446

\bibitem[{{Cordes} \& {Lazio}(2001)}]{cl01}
{Cordes}, J.~M., \& {Lazio}, T.~J.~W. 2001, \apj, 549, 997

\bibitem[{{Cordes} \& {Lazio}(2002)}]{cl02}
---. 2002, ArXiv Astrophysics e-prints: astro-ph/0207156

\bibitem[{{Cordes} \& {McLaughlin}(2003)}]{cm03}
{Cordes}, J.~M., \& {McLaughlin}, M.~A. 2003, \apj, 596, 1142

\bibitem[{{Cordes} \& {Rickett}(1998)}]{cr98}
{Cordes}, J.~M., \& {Rickett}, B.~J. 1998, \apj, 507, 846

\bibitem[{{Cordes} {et~al.}(2017){Cordes}, {Wasserman}, {Hessels}, {Lazio},
  {Chatterjee}, \& {Wharton}}]{cwh+17}
{Cordes}, J.~M., {Wasserman}, I., {Hessels}, J.~W.~T., {Lazio}, T.~J.~W.,
  {Chatterjee}, S., \& {Wharton}, R.~S. 2017, \apj, 842, 35

\bibitem[{{Cordes} {et~al.}(1985){Cordes}, {Weisberg}, \& {Boriakoff}}]{cwb85}
{Cordes}, J.~M., {Weisberg}, J.~M., \& {Boriakoff}, V. 1985, \apj, 288, 221

\bibitem[{{Desvignes} {et~al.}(2018){Desvignes}, {Eatough}, {Pen}, {Lee},
  {Mao}, {Karuppusamy}, {Schnitzeler}, {Falcke}, {Kramer}, {Wucknitz},
  {Spitler}, {Torne}, {Liu}, {Bower}, {Cognard}, {Lyne}, \&
  {Stappers}}]{dep+18}
{Desvignes}, G., {Eatough}, R.~P., {Pen}, U.~L., {Lee}, K.~J., {Mao}, S.~A.,
  {Karuppusamy}, R., {Schnitzeler}, D.~H.~F.~M., {Falcke}, H., {Kramer}, M.,
  {Wucknitz}, O., {Spitler}, L.~G., {Torne}, P., {Liu}, K., {Bower}, G.~C.,
  {Cognard}, I., {Lyne}, A.~G., \& {Stappers}, B.~W. 2018, \apjl, 852, L12

\bibitem[{{Farah} {et~al.}(2018){Farah}, {Flynn}, {Bailes}, {Jameson},
  {Bannister}, {Barr}, {Bateman}, {Bhandari}, {Caleb}, {Campbell-Wilson},
  {Chang}, {Deller}, {Green}, {Hunstead}, {Jankowski}, {Keane}, {Macquart},
  {M{\"o}ller}, {Onken}, {Os{\l}owski}, {Parthasarathy}, {Plant}, {Ravi},
  {Shannon}, {Tucker}, {Venkatraman Krishnan}, \& {Wolf}}]{ffb+18}
{Farah}, W., {Flynn}, C., {Bailes}, M., {Jameson}, A., {Bannister}, K.~W.,
  {Barr}, E.~D., {Bateman}, T., {Bhandari}, S., {Caleb}, M., {Campbell-Wilson},
  D., {Chang}, S.-W., {Deller}, A., {Green}, A.~J., {Hunstead}, R.,
  {Jankowski}, F., {Keane}, E., {Macquart}, J.-P., {M{\"o}ller}, A., {Onken},
  C.~A., {Os{\l}owski}, S., {Parthasarathy}, A., {Plant}, K., {Ravi}, V.,
  {Shannon}, R.~M., {Tucker}, B.~E., {Venkatraman Krishnan}, V., \& {Wolf}, C.
  2018, \mnras, 478, 1209

\bibitem[{{Gajjar} {et~al.}(2018){Gajjar}, {Siemion}, {Price}, {Law},
  {Michilli}, {Hessels}, {Chatterjee}, {Archibald}, {Bower}, {Brinkman},
  {Burke-Spolaor}, {Cordes}, {Croft}, {Enriquez}, {Foster}, {Gizani},
  {Hellbourg}, {Isaacson}, {Kaspi}, {Lazio}, {Lebofsky}, {Lynch}, {MacMahon},
  {McLaughlin}, {Ransom}, {Scholz}, {Seymour}, {Spitler}, {Tendulkar},
  {Werthimer}, \& {Zhang}}]{gsp+18}
{Gajjar}, V., {Siemion}, A.~P.~V., {Price}, D.~C., {Law}, C.~J., {Michilli},
  D., {Hessels}, J.~W.~T., {Chatterjee}, S., {Archibald}, A.~M., {Bower},
  G.~C., {Brinkman}, C., {Burke-Spolaor}, S., {Cordes}, J.~M., {Croft}, S.,
  {Enriquez}, J.~E., {Foster}, G., {Gizani}, N., {Hellbourg}, G., {Isaacson},
  H., {Kaspi}, V.~M., {Lazio}, T.~J.~W., {Lebofsky}, M., {Lynch}, R.~S.,
  {MacMahon}, D., {McLaughlin}, M.~A., {Ransom}, S.~M., {Scholz}, P.,
  {Seymour}, A., {Spitler}, L.~G., {Tendulkar}, S.~P., {Werthimer}, D., \&
  {Zhang}, Y.~G. 2018, ArXiv e-prints

\bibitem[{{Jankowski} {et~al.}(2018){Jankowski}, {van Straten}, {Keane},
  {Bailes}, {Barr}, {Johnston}, \& {Kerr}}]{jsk+18}
{Jankowski}, F., {van Straten}, W., {Keane}, E.~F., {Bailes}, M., {Barr},
  E.~D., {Johnston}, S., \& {Kerr}, M. 2018, \mnras, 473, 4436

\bibitem[{{Karuppusamy} {et~al.}(2010){Karuppusamy}, {Stappers}, \& {van
  Straten}}]{ksv10}
{Karuppusamy}, R., {Stappers}, B.~W., \& {van Straten}, W. 2010, \aap, 515, A36

\bibitem[{{Kramer} {et~al.}(2003){Kramer}, {Karastergiou}, {Gupta}, {Johnston},
  {Bhat}, \& {Lyne}}]{kkg+03}
{Kramer}, M., {Karastergiou}, A., {Gupta}, Y., {Johnston}, S., {Bhat},
  N.~D.~R., \& {Lyne}, A.~G. 2003, \aap, 407, 655

\bibitem[{{Law} {et~al.}(2017){Law}, {Abruzzo}, {Bassa}, {Bower},
  {Burke-Spolaor}, {Butler}, {Cantwell}, {Carey}, {Chatterjee}, {Cordes},
  {Demorest}, {Dowell}, {Fender}, {Gourdji}, {Grainge}, {Hessels}, {Hickish},
  {Kaspi}, {Lazio}, {McLaughlin}, {Michilli}, {Mooley}, {Perrott}, {Ransom},
  {Razavi-Ghods}, {Rupen}, {Scaife}, {Scott}, {Scholz}, {Seymour}, {Spitler},
  {Stovall}, {Tendulkar}, {Titterington}, {Wharton}, \& {Williams}}]{lab+17}
{Law}, C.~J., {Abruzzo}, M.~W., {Bassa}, C.~G., {Bower}, G.~C.,
  {Burke-Spolaor}, S., {Butler}, B.~J., {Cantwell}, T., {Carey}, S.~H.,
  {Chatterjee}, S., {Cordes}, J.~M., {Demorest}, P., {Dowell}, J., {Fender},
  R., {Gourdji}, K., {Grainge}, K., {Hessels}, J.~W.~T., {Hickish}, J.,
  {Kaspi}, V.~M., {Lazio}, T.~J.~W., {McLaughlin}, M.~A., {Michilli}, D.,
  {Mooley}, K., {Perrott}, Y.~C., {Ransom}, S.~M., {Razavi-Ghods}, N., {Rupen},
  M., {Scaife}, A., {Scott}, P., {Scholz}, P., {Seymour}, A., {Spitler}, L.~G.,
  {Stovall}, K., {Tendulkar}, S.~P., {Titterington}, D., {Wharton}, R.~S., \&
  {Williams}, P.~K.~G. 2017, \apj, 850, 76

\bibitem[{{Lazarus} {et~al.}(2015){Lazarus}, {Brazier}, {Hessels},
  {Karako-Argaman}, {Kaspi}, {Lynch}, {Madsen}, {Patel}, {Ransom}, {Scholz},
  {Swiggum}, {Zhu}, {Allen}, {Bogdanov}, {Camilo}, {Cardoso}, {Chatterjee},
  {Cordes}, {Crawford}, {Deneva}, {Ferdman}, {Freire}, {Jenet}, {Knispel},
  {Lee}, {van Leeuwen}, {Lorimer}, {Lyne}, {McLaughlin}, {Siemens}, {Spitler},
  {Stairs}, {Stovall}, \& {Venkataraman}}]{lbh+15}
{Lazarus}, P., {Brazier}, A., {Hessels}, J.~W.~T., {Karako-Argaman}, C.,
  {Kaspi}, V.~M., {Lynch}, R., {Madsen}, E., {Patel}, C., {Ransom}, S.~M.,
  {Scholz}, P., {Swiggum}, J., {Zhu}, W.~W., {Allen}, B., {Bogdanov}, S.,
  {Camilo}, F., {Cardoso}, F., {Chatterjee}, S., {Cordes}, J.~M., {Crawford},
  F., {Deneva}, J.~S., {Ferdman}, R., {Freire}, P.~C.~C., {Jenet}, F.~A.,
  {Knispel}, B., {Lee}, K.~J., {van Leeuwen}, J., {Lorimer}, D.~R., {Lyne},
  A.~G., {McLaughlin}, M.~A., {Siemens}, X., {Spitler}, L.~G., {Stairs}, I.~H.,
  {Stovall}, K., \& {Venkataraman}, A. 2015, \apj, 812, 81

\bibitem[{{Lazarus} {et~al.}(2016){Lazarus}, {Karuppusamy}, {Graikou},
  {Caballero}, {Champion}, {Lee}, {Verbiest}, \& {Kramer}}]{lkg+16}
{Lazarus}, P., {Karuppusamy}, R., {Graikou}, E., {Caballero}, R.~N.,
  {Champion}, D.~J., {Lee}, K.~J., {Verbiest}, J.~P.~W., \& {Kramer}, M. 2016,
  \mnras, 458, 868

\bibitem[{{Lorimer} {et~al.}(2007){Lorimer}, {Bailes}, {McLaughlin},
  {Narkevic}, \& {Crawford}}]{lbm+07}
{Lorimer}, D.~R., {Bailes}, M., {McLaughlin}, M.~A., {Narkevic}, D.~J., \&
  {Crawford}, F. 2007, Science, 318, 777

\bibitem[{{Lundgren} {et~al.}(1995){Lundgren}, {Cordes}, {Ulmer}, {Matz},
  {Lomatch}, {Foster}, \& {Hankins}}]{lcu+95}
{Lundgren}, S.~C., {Cordes}, J.~M., {Ulmer}, M., {Matz}, S.~M., {Lomatch}, S.,
  {Foster}, R.~S., \& {Hankins}, T. 1995, \apj, 453, 433

\bibitem[{{Main} {et~al.}(2018){Main}, {Yang}, {Chan}, {Li}, {Lin}, {Mahajan},
  {Pen}, {Vanderlinde}, \& {van Kerkwijk}}]{myc+18}
{Main}, R., {Yang}, I., {Chan}, V., {Li}, D., {Lin}, F.~X., {Mahajan}, N.,
  {Pen}, U.-L., {Vanderlinde}, K., \& {van Kerkwijk}, M.~H. 2018, \nat, 557,
  522

\bibitem[{{Marcote} {et~al.}(2017){Marcote}, {Paragi}, {Hessels}, {Keimpema},
  {van Langevelde}, {Huang}, {Bassa}, {Bogdanov}, {Bower}, {Burke-Spolaor},
  {Butler}, {Campbell}, {Chatterjee}, {Cordes}, {Demorest}, {Garrett}, {Ghosh},
  {Kaspi}, {Law}, {Lazio}, {McLaughlin}, {Ransom}, {Salter}, {Scholz},
  {Seymour}, {Siemion}, {Spitler}, {Tendulkar}, \& {Wharton}}]{mph+17}
{Marcote}, B., {Paragi}, Z., {Hessels}, J.~W.~T., {Keimpema}, A., {van
  Langevelde}, H.~J., {Huang}, Y., {Bassa}, C.~G., {Bogdanov}, S., {Bower},
  G.~C., {Burke-Spolaor}, S., {Butler}, B.~J., {Campbell}, R.~M., {Chatterjee},
  S., {Cordes}, J.~M., {Demorest}, P., {Garrett}, M.~A., {Ghosh}, T., {Kaspi},
  V.~M., {Law}, C.~J., {Lazio}, T.~J.~W., {McLaughlin}, M.~A., {Ransom}, S.~M.,
  {Salter}, C.~J., {Scholz}, P., {Seymour}, A., {Siemion}, A., {Spitler},
  L.~G., {Tendulkar}, S.~P., \& {Wharton}, R.~S. 2017, \apjl, 834, L8

\bibitem[{{Metzger} {et~al.}(2017){Metzger}, {Berger}, \& {Margalit}}]{mbm17}
{Metzger}, B.~D., {Berger}, E., \& {Margalit}, B. 2017, \apj, 841, 14

\bibitem[{{Michilli} {et~al.}(2018){Michilli}, {Seymour}, {Hessels}, {Spitler},
  {Gajjar}, {Archibald}, {Bower}, {Chatterjee}, {Cordes}, {Gourdji}, {Heald},
  {Kaspi}, {Law}, {Sobey}, {Adams}, {Bassa}, {Bogdanov}, {Brinkman},
  {Demorest}, {Fernandez}, {Hellbourg}, {Lazio}, {Lynch}, {Maddox}, {Marcote},
  {McLaughlin}, {Paragi}, {Ransom}, {Scholz}, {Siemion}, {Tendulkar}, {van
  Rooy}, {Wharton}, \& {Whitlow}}]{msh+18}
{Michilli}, D., {Seymour}, A., {Hessels}, J.~W.~T., {Spitler}, L.~G., {Gajjar},
  V., {Archibald}, A.~M., {Bower}, G.~C., {Chatterjee}, S., {Cordes}, J.~M.,
  {Gourdji}, K., {Heald}, G.~H., {Kaspi}, V.~M., {Law}, C.~J., {Sobey}, C.,
  {Adams}, E.~A.~K., {Bassa}, C.~G., {Bogdanov}, S., {Brinkman}, C.,
  {Demorest}, P., {Fernandez}, F., {Hellbourg}, G., {Lazio}, T.~J.~W., {Lynch},
  R.~S., {Maddox}, N., {Marcote}, B., {McLaughlin}, M.~A., {Paragi}, Z.,
  {Ransom}, S.~M., {Scholz}, P., {Siemion}, A.~P.~V., {Tendulkar}, S.~P., {van
  Rooy}, P., {Wharton}, R.~S., \& {Whitlow}, D. 2018, \nat, 553, 182

\bibitem[{{Mikami} {et~al.}(2016){Mikami}, {Asano}, {Tanaka}, {Kisaka},
  {Sekido}, {Takefuji}, {Takeuchi}, {Misawa}, {Tsuchiya}, {Kita}, {Yonekura},
  \& {Terasawa}}]{mat+16}
{Mikami}, R., {Asano}, K., {Tanaka}, S.~J., {Kisaka}, S., {Sekido}, M.,
  {Takefuji}, K., {Takeuchi}, H., {Misawa}, H., {Tsuchiya}, F., {Kita}, H.,
  {Yonekura}, Y., \& {Terasawa}, T. 2016, \apj, 832, 212

\bibitem[{{Oostrum} {et~al.}(2017){Oostrum}, {van Leeuwen}, {Attema}, {van
  Cappellen}, {Connor}, {Hut}, {Maan}, {Oosterloo}, {Petroff}, {van der
  Schuur}, {Sclocco}, \& {Verheijen}}]{ola+17}
{Oostrum}, L.~C., {van Leeuwen}, J., {Attema}, J., {van Cappellen}, W.,
  {Connor}, L., {Hut}, B., {Maan}, Y., {Oosterloo}, T.~A., {Petroff}, E., {van
  der Schuur}, D., {Sclocco}, A., \& {Verheijen}, M.~A.~W. 2017, The
  Astronomer's Telegram, 10693

\bibitem[{{Oppermann} {et~al.}(2018){Oppermann}, {Yu}, \& {Pen}}]{oyp18}
{Oppermann}, N., {Yu}, H.-R., \& {Pen}, U.-L. 2018, \mnras, 475, 5109

\bibitem[{{Petroff} {et~al.}(2016){Petroff}, {Barr}, {Jameson}, {Keane},
  {Bailes}, {Kramer}, {Morello}, {Tabbara}, \& {van Straten}}]{pbj+16}
{Petroff}, E., {Barr}, E.~D., {Jameson}, A., {Keane}, E.~F., {Bailes}, M.,
  {Kramer}, M., {Morello}, V., {Tabbara}, D., \& {van Straten}, W. 2016, \pasa,
  33, e045

\bibitem[{{Ransom}(2001)}]{r01}
{Ransom}, S.~M. 2001, PhD thesis, Harvard University

\bibitem[{{Scholz} {et~al.}(2017){Scholz}, {Bogdanov}, {Hessels}, {Lynch}, {\bf
  {Spitler}, L.~G.}, {Bassa}, {Bower}, {Burke-Spolaor}, {Butler}, {Chatterjee},
  {Cordes}, {Gourdji}, {Kaspi}, {Law}, {Marcote}, {McLaughlin}, {Michilli},
  {Paragi}, {Ransom}, {Seymour}, {Tendulkar}, \& {Wharton}}]{sbh+17}
{Scholz}, P., {Bogdanov}, S., {Hessels}, J.~W.~T., {Lynch}, R.~S., {\bf
  {Spitler}, L.~G.}, {Bassa}, C.~G., {Bower}, G.~C., {Burke-Spolaor}, S.,
  {Butler}, B.~J., {Chatterjee}, S., {Cordes}, J.~M., {Gourdji}, K., {Kaspi},
  V.~M., {Law}, C.~J., {Marcote}, B., {McLaughlin}, M.~A., {Michilli}, D.,
  {Paragi}, Z., {Ransom}, S.~M., {Seymour}, A., {Tendulkar}, S.~P., \&
  {Wharton}, R.~S. 2017, \apj, 846, 80

\bibitem[{{Scholz} {et~al.}(2016){Scholz}, {Spitler}, {Hessels}, {Chatterjee},
  {Cordes}, {Kaspi}, {Wharton}, {Bassa}, {Bogdanov}, {Camilo}, {Crawford},
  {Deneva}, {van Leeuwen}, {Lynch}, {Madsen}, {McLaughlin}, {Mickaliger},
  {Parent}, {Patel}, {Ransom}, {Seymour}, {Stairs}, {Stappers}, \&
  {Tendulkar}}]{ssh+16a}
{Scholz}, P., {Spitler}, L.~G., {Hessels}, J.~W.~T., {Chatterjee}, S.,
  {Cordes}, J.~M., {Kaspi}, V.~M., {Wharton}, R.~S., {Bassa}, C.~G.,
  {Bogdanov}, S., {Camilo}, F., {Crawford}, F., {Deneva}, J., {van Leeuwen},
  J., {Lynch}, R., {Madsen}, E.~C., {McLaughlin}, M.~A., {Mickaliger}, M.,
  {Parent}, E., {Patel}, C., {Ransom}, S.~M., {Seymour}, A., {Stairs}, I.~H.,
  {Stappers}, B.~W., \& {Tendulkar}, S.~P. 2016, \apj, 833, 177

\bibitem[{{Spitler} {et~al.}(2012){Spitler}, {Cordes}, {Chatterjee}, \&
  {Stone}}]{sccs12}
{Spitler}, L.~G., {Cordes}, J.~M., {Chatterjee}, S., \& {Stone}, J. 2012, \apj,
  748, 73

\bibitem[{{Spitler} {et~al.}(2014){Spitler}, {Cordes}, {Hessels}, {Lorimer},
  {McLaughlin}, {Chatterjee}, {Crawford}, {Deneva}, {Kaspi}, {Wharton},
  {Allen}, {Bogdanov}, {Brazier}, {Camilo}, {Freire}, {Jenet},
  {Karako-Argaman}, {Knispel}, {Lazarus}, {Lee}, {van Leeuwen}, {Lynch},
  {Ransom}, {Scholz}, {Siemens}, {Stairs}, {Stovall}, {Swiggum},
  {Venkataraman}, {Zhu}, {Aulbert}, \& {Fehrmann}}]{sch+14}
{Spitler}, L.~G., {Cordes}, J.~M., {Hessels}, J.~W.~T., {Lorimer}, D.~R.,
  {McLaughlin}, M.~A., {Chatterjee}, S., {Crawford}, F., {Deneva}, J.~S.,
  {Kaspi}, V.~M., {Wharton}, R.~S., {Allen}, B., {Bogdanov}, S., {Brazier}, A.,
  {Camilo}, F., {Freire}, P.~C.~C., {Jenet}, F.~A., {Karako-Argaman}, C.,
  {Knispel}, B., {Lazarus}, P., {Lee}, K.~J., {van Leeuwen}, J., {Lynch}, R.,
  {Ransom}, S.~M., {Scholz}, P., {Siemens}, X., {Stairs}, I.~H., {Stovall}, K.,
  {Swiggum}, J.~K., {Venkataraman}, A., {Zhu}, W.~W., {Aulbert}, C., \&
  {Fehrmann}, H. 2014, \apj, 790, 101

\bibitem[{{Spitler} {et~al.}(2016){Spitler}, {Scholz}, {Hessels}, {Bogdanov},
  {Brazier}, {Camilo}, {Chatterjee}, {Cordes}, {Crawford}, {Deneva}, {Ferdman},
  {Freire}, {Kaspi}, {Lazarus}, {Lynch}, {Madsen}, {McLaughlin}, {Patel},
  {Ransom}, {Seymour}, {Stairs}, {Stappers}, {van Leeuwen}, \& {Zhu}}]{ssh+16}
{Spitler}, L.~G., {Scholz}, P., {Hessels}, J.~W.~T., {Bogdanov}, S., {Brazier},
  A., {Camilo}, F., {Chatterjee}, S., {Cordes}, J.~M., {Crawford}, F.,
  {Deneva}, J., {Ferdman}, R.~D., {Freire}, P.~C.~C., {Kaspi}, V.~M.,
  {Lazarus}, P., {Lynch}, R., {Madsen}, E.~C., {McLaughlin}, M.~A., {Patel},
  C., {Ransom}, S.~M., {Seymour}, A., {Stairs}, I.~H., {Stappers}, B.~W., {van
  Leeuwen}, J., \& {Zhu}, W.~W. 2016, \nat, 531, 202

\bibitem[{{Tendulkar} {et~al.}(2017){Tendulkar}, {Bassa}, {Cordes}, {Bower},
  {Law}, {Chatterjee}, {Adams}, {Bogdanov}, {Burke-Spolaor}, {Butler},
  {Demorest}, {Hessels}, {Kaspi}, {Lazio}, {Maddox}, {Marcote}, {McLaughlin},
  {Paragi}, {Ransom}, {Scholz}, {Seymour}, {Spitler}, {van Langevelde}, \&
  {Wharton}}]{tbc+17}
{Tendulkar}, S.~P., {Bassa}, C.~G., {Cordes}, J.~M., {Bower}, G.~C., {Law},
  C.~J., {Chatterjee}, S., {Adams}, E.~A.~K., {Bogdanov}, S., {Burke-Spolaor},
  S., {Butler}, B.~J., {Demorest}, P., {Hessels}, J.~W.~T., {Kaspi}, V.~M.,
  {Lazio}, T.~J.~W., {Maddox}, N., {Marcote}, B., {McLaughlin}, M.~A.,
  {Paragi}, Z., {Ransom}, S.~M., {Scholz}, P., {Seymour}, A., {Spitler}, L.~G.,
  {van Langevelde}, H.~J., \& {Wharton}, R.~S. 2017, \apjl, 834, L7

\bibitem[{{Thornton} {et~al.}(2013){Thornton}, {Stappers}, {Bailes},
  {Barsdell}, {Bates}, {Bhat}, {Burgay}, {Burke-Spolaor}, {Champion}, {Coster},
  {D'Amico}, {Jameson}, {Johnston}, {Keith}, {Kramer}, {Levin}, {Milia}, {Ng},
  {Possenti}, \& {van Straten}}]{tsb+13}
{Thornton}, D., {Stappers}, B., {Bailes}, M., {Barsdell}, B., {Bates}, S.,
  {Bhat}, N.~D.~R., {Burgay}, M., {Burke-Spolaor}, S., {Champion}, D.~J.,
  {Coster}, P., {D'Amico}, N., {Jameson}, A., {Johnston}, S., {Keith}, M.,
  {Kramer}, M., {Levin}, L., {Milia}, S., {Ng}, C., {Possenti}, A., \& {van
  Straten}, W. 2013, Science, 341, 53

\bibitem[{{Wielebinski}(2007)}]{w07}
{Wielebinski}, R. 2007, Astronomische Nachrichten, 328, 388

\end{thebibliography}

\end{document}